# Size-Change Termination as a Contract

## Dynamically and Statically Enforcing Termination for Higher-Order Programs

Phúc C. Nguyễn
University of Maryland
College Park, MD, USA

Thomas Gilray
University of Alabama
Birmingham, AL, USA

Sam Tobin-Hochstadt
Indiana University
Bloomington, IN, USA

David Van Horn
University of Maryland
College Park, MD, USA

## Abstract

Termination is an important but undecidable program property, which has led to a large body of work on static methods for conservatively predicting or enforcing termination. One such method is the *size-change termination* approach of Lee, Jones, and Ben-Amram, which operates in two phases: (1) abstract programs into "size-change graphs," and (2) check these graphs for the *size-change property*: the existence of paths that lead to infinite decreasing sequences.

We transpose these two phases with an operational semantics that accounts for the *run-time enforcement of the size-change property*, postponing (or entirely avoiding) program abstraction. This choice has two key consequences: (1) size-change termination can be checked at run-time and (2) termination can be rephrased as a safety property analyzed using existing methods for systematic abstraction.

We formulate run-time size-change checks as *contracts* in the style of Findler and Felleisen. The result compliments existing contracts that enforce partial correctness specifications to obtain *contracts for total correctness*. Our approach combines the robustness of the size-change principle for termination with the precise information available at run-time. It has tunable overhead and can check for nontermination without the conservativeness necessary in static checking. To obtain a sound and computable termination analysis, we apply existing abstract interpretation techniques directly to the operational semantics, avoiding the need for custom abstractions for termination. The resulting analyzer is competitive with with existing, purpose-built analyzers.

## 1 Size-change contracts

*A fool's errand*  Imagine for a moment there existed a run-time mechanism for checking whether a program, in its current state, will run forever or eventually terminate. Such a check would be eminently useful. Any run-time mechanism for enforcing partial correctness could easily be made to enforce total correctness by use of this check. Moreover, static verification of termination would boil down to proving these run-time checks always succeed, much like how type systems prove run-time tag checks always succeed.

Of course, whether a program eventually terminates is one of the most useful, yet fundamentally and famously unknowable, properties of programs [16, 48]. Moreover, due to its nature as a liveness property—it cannot be violated in a finite execution—it cannot be directly checked at run-time.

*An indirect tack*  Despite this situation, an indirect partial solution is possible by instead considering a safety property that implies the liveness property. This indirect approach underlies successful static termination analysis tools such as Terminator [14]. Given such a safety property, enforcing it at run-time would ensure a nonterminating program would eventually "go wrong" by violating the safety property, at which point it could be stopped. The one, unavoidable, wrinkle is that there will be some programs that run astray of the safety property, despite eventually terminating. In this approach, static verification of termination could, as suggested before, be phrased and designed just as any other safety verification problem by proving the impossibility of a run-time check failure. This approach has the added advantage that any program can be dynamically monitored regardless of whether it can be statically verified.

***A universal safety property for termination*** To design a run-time termination checker, the critical question is: what is a good safety property to enforce that implies termination? Tools such as Terminator, AProVE, and many others discover a program-specific termination argument, either through static analysis or CEGAR-style refinement. While such approaches have proved quite successful in learning complex termination arguments, these approaches undermine the ability to dynamically monitor termination.

To remedy the situation, we propose using a universal property. A promising candidate is the so-called *size-change principle* of Lee et al. [29]. The principle has proved useful in static termination checking and has a well understood theory. Unfortunately, the original work on size-change termination, which was developed for static verification, defines the size-change principle as a property of a program *abstraction*: a set of so-called size-change graphs (roughly a program call graph annotated with information about non-ascending data flows between function parameters).

***This paper*** We propose a run-time check inspired by the size-change principle for program termination that *dynamically* builds and checks precise size-change graphs. This dynamic mechanism is useful in its own right, but also can be used as a basis for designing static termination checkers. Such static checkers can benefit from advances in static analysis, particularly in abstract interpretation, since termination checks are integrated into the language specification and do not require custom abstractions or algorithms.

We formalize a semantics for a higher-order functional language that enforces the size-change principle, thereby ensuring all programs terminate (§3). Moreover, we introduce a behavioral software contract, in the style of Findler and Felleisen [17], that enables the selective enforcement of size-change termination. Such contracts, when combined with traditional pre- and post-condition contracts, form a notion of contracts for total-correctness.

We also develop a static termination checker (§4) by applying the static contract verification technique of Nguyễn et al. [34] to the size-change semantics. The resulting tool has no termination analysis specific abstractions, it simply treats the size-change principle check as it would any run-time check, and yet an empirical evaluation (§5) shows that it is competitive with several state-of-the-art purpose-built termination analyzers: Liquid Haskell, Isabelle, and ACL2.

***Contributions*** This paper contributes:

1. a semantic account of the size-change principle,
2. a proved-correct contract for size-change-based termination of functions,
3. an implementation technique that preserves proper tail-calls and enables tunable run-time overhead, and
4. a static termination checker obtained by generic abstract interpretation techniques.

## 2 Examples and intuitions

This section develops intuitions for how dynamic checking of *size-change termination* (SCT) works via worked examples. We begin by sketching how SCT works in the original static setting of Lee et al. [29].

### 2.1 The factorial of termination papers

Consider the Ackermann function, the standard-bearer of examples for papers on termination due its simplicity as a total—but not primitive recursive—function, presented here in Scheme notation:

```
1  (define (ack m n)
2    (cond [(= 0 m) (+ 1 n)]
3          [(= 0 n) (ack (- m 1) 1)]
4          [else   (ack (- m 1)
5                       (ack m (- n 1)))]))
```

For the moment, assume the function is only applied to natural numbers. Under that assumption, ack always terminates and the SCT method suffices to prove it.

***Safe size-change graphs:*** The approach starts by using program analysis or abstract interpretation to enumerate the ways in which a call to ack could result in a subsequent call to ack before returning. We can see there are three potential recursive calls within the function definition on lines 3, 4, and 5. For each of these calls, describe the pairwise relations between the arguments of the call and recursive call in terms of their size. (The original SCT approach assumes the language has only well-founded data types with a known partial order.)

So for example, consider the possible call:

$$(\text{ack m n}) \rightsquigarrow (\text{ack (- m 1) 1}).$$

There are two parameters, so we consider four possible size-change relations between the inputs and recursive call. It is clear that the m parameter is strictly smaller in the recursive call compared to the input of the original call. This change is described with a "size-change graph," $\{(\text{m} \downarrow \text{m})\}$, which is a binary relation saying that whatever value is given for m in the original call will become a strictly smaller argument m in the recursive call. But there is no size-change relation between the original input n and recursive parameter m or n, nor between the original m and recursive n, which we know is 1: each could become larger, smaller, or stay the same.

Moving on to the call in line 5:

$$(\text{ack m n}) \rightsquigarrow (\text{ack m (- n 1)}),$$

we can see that m is unchanged and n is strictly smaller between calls (but there's no relation between m and n), so we describe this call with the graph: $\{(\text{m} \;\overline{\downarrow}\; \text{m}), (\text{n} \downarrow \text{n})\}$, which says m is non-ascending and n is descending.

Finally, consider the call in line 4:

$$(\text{ack m n}) \rightsquigarrow (\text{ack (- m 1) ... )},$$

where the elided code is the nested call to ack of line 5. Here it is clear that m strictly descends, but unclear what happens with n. So we can describe this call with the size-change graph as used for the call in line 3.

At this point, we now have a sound collection of size-change graphs for all possible successive calls to ack. They are sound since they properly account for all possible strict descent or non-ascending transitions that occur in recursive calls at run-time. As a side note: it is always safe to omit graph arcs (potentially losing sufficient evidence to prove termination), but all arcs included in a graph must soundly over-approximate all possible run-time behaviors.

***Size-change principle:*** The next task is to check this set of graphs for the *size-change termination principle* (SCP) to see if every infinite computation would give rise to an infinitely decreasing value sequence, according to the size-change graphs. To do this, we consider closing the set of graphs under sequential composition of size-change graphs. The sequential composition of two graphs models two successive calls to construct the size-change from the first to last call, and is defined, informally, as follows: there is a strict descending arc between two parameters, if there exists a path between the parameters containing a strict descent; there is a is a non-ascending arc if there exists a path containing only non-ascent arcs. Otherwise, there is no path.

Coincidentally, the set of graphs for ack is already closed under sequential composition, but to see an example, here's the sequential composition of calling ack on line 3 (or 4) followed by ack on line 5:

$$\{(m \downarrow m)\}; \{(m \sqsubset^{=} m), (n \downarrow n)\} = \{(m \downarrow m)\},$$

which is equivalent, in terms of size-change, as calling ack on line 3 (or 4).

Once closed, we check each size-change graph to see if it

1. is idempotent, i.e. $g; g = g$, and
2. lacks a self descending arc, i.e. $(x \downarrow x)$ for some parameter $x$.

If such a graph exists, it represents a potential sequence of calls that can be iterated infinitely often with no descent and thus violates the size-change principle. If it lacks such a graph, the program terminates. In the case of ack both graphs have self-descending arcs and therefore terminates.

***Dynamic size-change graphs:*** Having established the basic notions of the static SCT approach, we now turn to a dynamic approach to monitoring size-change termination.

The main idea is that rather than rely upon a program analysis to enumerate the various ways a function may call itself, we simply run the program and observe such calls. Each time a function invokes itself, a size-change graph is dynamically generated. Throughout a computation, the call sequence of size-change graphs is accumulated. Before entering a function call, the current call sequence is checked for the size-change principle. If it is violated, the program is stopped and an error signalled; otherwise the call proceeds.

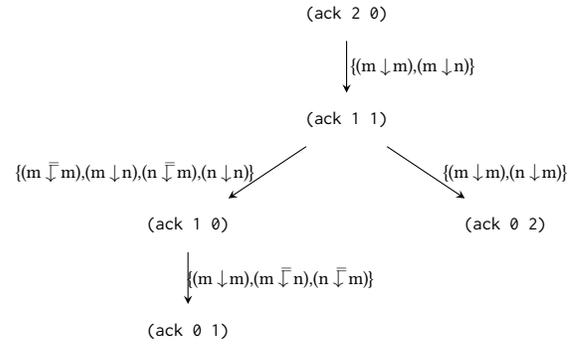

**Figure 1.** Calls and size changes for (ack 2 0)

A program violating the size-change principle eventually accumulates a call sequence witnessing the violation; a program maintaining the principle eventually terminates.

In a similar vein, we need not rely on static analysis to infer the size-change relation between arguments. At run-time, there are concrete values available at both the call and recursive call site. Inferring the size-change graph boils down to checking a partial order pairwise on the arguments. This is both easy to do and potentially much more precise than the static approach. For example, there may be size-change relations that hold on the particular path of execution under scrunity, which do not hold in general.

To make things concrete, reconsider ack. When switching perspectives to the dynamic setting, we are no longer concerned with proving termination for all possible executions of the function, but rather with a particular application. Consider (ack 2 0). The complete tree of call sequences and generated size-change graphs is shown in Figure 1, but let us step through its construction. In calling (ack 2 0), control reaches the recursive call on line 3, so we have the call sequence:

$$(\text{ack } 2 \ 0) \rightsquigarrow (\text{ack } 1 \ 1),$$

from which we can read off the size-change graph. Just as in the static case, we have $(m \downarrow m)$, but additionally, we know that $(m \downarrow n)$. This fact does not hold in all runs of ack, but it holds in this one.

*Aside:* it is worth noting that this additional program fact is not necessary in this particular example. After all, we have statically proven ack terminates in all cases using less information. But for the purposes of illustration, we can see that more information is available at run-time; and in principle, it is possible to safely execute size-change terminating programs that are not statically verifiable, just as by analogy it is possible to dynamically monitor type safety of programs that do not trigger run-time type errors, yet are statically ill-typed.

Returning to the example: having generated the graph for this call, we then check the SCT principle for the active sequence of calls; in this case there is just the one graph: $\{(m \downarrow m), (m \downarrow n)\}$, which satisfies the size-change property, so execution proceeds.

Now (ack 1 1) reaches the `else` branch and invokes a recursive call to (ack 1 0) on line 5. This call generates the graph $\{(m \overline{\downarrow} m), (m \downarrow n), (n \overline{\downarrow} m), (n \overline{\downarrow} n)\}$. We now check the size-change graphs of the sequence leading to this point, i.e., the size-change graphs of:

$$(\text{ack } 2 \ 0) \rightsquigarrow (\text{ack } 1 \ 1) \rightsquigarrow (\text{ack } 1 \ 0),$$

and determine if the size-change property holds, which it does. Now (ack 1 0) reaches (ack 0 1) with graph $\{(m \downarrow n), (n \downarrow m), (m \overline{\downarrow} m), (n \overline{\downarrow} n)\}$, and the call sequence still satisfies SCP. At this point (ack 0 1) terminates with 2. This brings control back to the evaluation of (ack 1 1), which is now ready to proceed to second call to ack on line 4 with the arguments (ack 0 1). At this point, we have the call sequence:

$$(\text{ack } 2 \ 0) \rightsquigarrow (\text{ack } 1 \ 1) \rightsquigarrow (\text{ack } 0 \ 2).$$

Note the calls to (ack 1 0) and (ack 0 1) are no longer active since they have returned. Again we check the SCP of the size-change graph sequence for active calls, which holds and the program terminates.

***A sometimes-buggy Ackermann:*** We have seen how runtime SCT monitoring works for programs that maintain the size-change principle, but what about buggy programs that do not? Consider the ack example, but change the call on line 4 from (ack (- m 1) ...) to (ack m ...). Computing (ack 2 0) would proceed as before until reaching the call on line 5, corresponding to the right branch of the tree in Figure 1, i.e. representing the call sequence:

$$(\text{ack } 2 \ 0) \rightsquigarrow (\text{ack } 1 \ 1) \rightsquigarrow (\text{ack } 1 \ 2),$$

whose last size-change graph is now $\{(m \overline{\downarrow} m), (n \overline{\downarrow} m)\}$. But this graph is idempotent and contains no self-descents, so at the point of this call a size-change violation is signaled.

## 2.2 Keeping closures in order

The original formulation of SCT was for a first-order functional language with a well-founded partial order on values. This was done largely to simplify the first phase of static SCT verification where call-graphs and size-change relations are generated. In higher-order languages, however, computing call-graphs is itself a significant, extensively studied problem [32]. In the dynamic formulation, higher-order functions do not pose a serious challenge since calls are observed as they occur.

The one remaining issue concerns the choice of partial order for functions. We make a simple choice and consider all closures to be incomparable. Consequently, no termination proof goes through by an argument about closure size. This is not to say that all programs that use higher-order functions will be rejected by the size-change monitor, just that they must have some descent on base values between calls to the same function. Our empirical evaluation (§5) confirms this is a reasonable choice. To illustrate, let us consider a program that recursively accumulates a closure and eventually applies it in the base case of the function.

Consider a len function for lists, written in CPS:

```
1  (define (len l) (loop l (λ (x) x)))
2  (define (loop l k)
3    (cond [(empty? l) (k 0)]
4          [(cons? l)
5           (loop (rest l) (λ (n) (k (+ 1 n))))]))
```

Static analysis of size-change termination relies on an underlying control-flow graph, which must eventually conflate all closures generated on line 5, regardless of call-sensitivity. This results in a spurious loop where each closure bound to k may appear to call one with a *larger* argument, failing the size-change principle.

Dynamic checking of size-change termination does not have this problem, because all the closures are exact and distinct. Even though the number of closures is arbitrary, they are finite up to the previous loop descending on l, which has been proven to terminate. The call sequence for (len '(2 1)), which is a sequence of tail-calls:

$$(\text{len } '(2 \ 1)) \rightsquigarrow (\text{loop } '(2 \ 1) \ (\lambda \ (x) \ x)) \rightsquigarrow$$
$$(\text{loop } '(1) \ k_1) \rightsquigarrow (\text{loop } '() \ k_2) \rightsquigarrow$$
$$(k_2 \ 0) \rightsquigarrow (k_1 \ 1) \rightsquigarrow ((\lambda \ (x) \ x) \ 2) \rightsquigarrow 2$$

The recursive calls of loop to itself are easily proven safe through descent on the list. The successive calls to continuations are arbitrarily many but finite. Here $k_1$ and $k_2$ stand for different closures of the $(\lambda \ (n) \ (k \ (+ \ 1 \ n)))$ term. The computation proceeds to an answer since SCP is only checked between calls to the *same* closure, directly or indirectly.

It is possible to define a partial order on closures, and this may be a worthwhile addition to our approach. For example, Jones and Bohr [25] extend SCT to the untyped λ-calculus and use a partial order based on closure depth to order functions. In theory, this could be used to dynamically order closures in our approach, too, however pragmatically, it requires run-time facilities for "opening" closures [43], which are not typically available.

## 2.3 Termination and blame

It is useful to assert size-change termination of particular functions, without necessarily asserting termination of the whole program. For this reason, we introduce a contract, terminating/c, in the style of Findler and Felleisen [17]. One key component of contract semantics is *blame* to explain the party at fault in contract errors. While our formal model does not represent blame, our implementation does. The addition of blame is at once simple and powerful in the setting of termination contracts. Each terminating/c use marks a blame party, and if the function so wrapped fails to terminate on some call, that location in the program is blamed. No sophisticated run-time machinery is required.

The addition of blame enables a virtuous cycle in program development. If a terminating function f calls g, then any failure to terminate on the part of g will be blamed on f. To protect themselves from being blamed, the author of f can in turn impose the same contract on g, leading to richer specifications and precise errors pinpointing the faulty component. Finally, the provision of size-change termination contracts enables a gradual-typing-style integration of total and partial program components.

## 2.4 The power of dynamic enforcement

Checking termination of a interpreter for a language that is Turing-complete is challenging—after all, the interpreter does not terminate on all programs. Nevertheless, dynamic size-change monitoring allows the interpretation of many interesting programs to finish. In Figure 2, we present a λ-calculus implementation that first compiles the term to a procedure and then applies this procedure to an environment. The compilation itself terminates by structural recursion, which is simple to check, but the compilation result is a procedure whose termination is not obvious. In fact, in this example, the first test program $c_1$ terminates when run, but $c_2$ loops infinitely. Dynamic size-change monitoring flexibly allows the first one to finish, and quickly catches the divergence in the second one. The ability to check for termination of specialized programs highlights the advantages of dynamic termination checking.

Execution of ($c_1$ (hash)) terminates because no function ever calls itself with a non-decreasing argument. In contrast, during the execution of ($c_2$ (hash)), the compilation result of (λ (y) (y y)) calls itself (indirectly) with a non-decreasing argument (in this case, identical), hence caught by the monitoring. As shown in the evaluation section (§5), our implementation is able to confirm the termination of a Scheme interpreter executing merge-sort.

## 3 Dynamic SCT monitoring

This section introduces language $λ_{SCT}$, which is λ-calculus, extended with base values and primitive operations, and with a modified semantics ensuring that all programs terminate. Figure 3 shows $λ_{SCT}$'s syntax and semantics.

```
1  (define comp
2    (terminating/c
3      (λ (e)
4        (match e
5          [`(λ (,x) ,e)
6           (let ([c (comp e)])
7             (λ (ρ) (λ (z) (c (hash-set ρ x z)))))]
8          [`(,e₁ ,e₂)
9           (let ([c₁ (comp e₁)] [c₂ (comp e₂)])
10            (λ (ρ) ((c₁ ρ) (c₂ ρ))))]
11         [(? symbol? x) (λ (ρ) (hash-ref ρ x))]))))
12 (define c₁
13   (terminating/c ; Okay
14     (comp '((λ (x) (x x)) (λ (y) y)))))
15 (define c₂
16   (terminating/c ; Okay
17     (comp '((λ (x) (x x)) (λ (y) (y y))))))
18 (c₁ (hash)) ; Okay
19 (c₂ (hash)) ; Error
```

**Figure 2.** A checked λ-calculus implementation

## 3.1 A terminating semantics

The domain of values ($v$) in $λ_{SCT}$ includes primitives ($o$), integers ($n$), pairs ($v_1, v_2$), and closures (($\vec{x}, e, ρ$)). No primitive in $λ_{SCT}$ is allowed to cause divergence.

We present the semantics of $λ_{SCT}$ in Figure 3. The semantics is defined by relation $ρ, m ∪ \{⊥\} ⊢ e ⬇ α$, which extends the standard semantics by accumulating a size-change table $m$. The size-change table maps each function ($v$) to the most recent arguments it was applied to, in the current dynamic extent, as well as a sequence of size-change graphs ($\vec{g}$) recording ways in which arguments of ($v$) descend. A size-change graph ($g$) is a set of arcs of the form ($i ↓ j$) or ($i \stackrel{↓}{=} j$), indicating that the $i$-th argument always strictly descends ($↓$) or never ascends ($\stackrel{↓}{=}$) to the $j$-th argument.

An evaluation answer ($α$) can be a value, run-time error (error[RT]), or size-change error (error[SC]). A run-time error is one resulting from misuse of language constructs as standard in a programming language (e.g. applying a primitive to arguments not in its intended domain, applying a non-function, or a function of the wrong arity, etc.). A size-change error is one raised by size-change monitoring upon detecting a size-change violation. We omit rules that introduce run-time errors and error propagation, as they are entirely standard and not the focus of this paper.

Rule *[SC-App-Clo]* shows application of a closure. In $λ_{SCT}$, all applications are enforced to have the size-change property. Before executing the function's body as in the standard semantics, we update the size-change table and guard against a violation to the size-change property. Helper function *upd* updates the size-change table with the function's

$$
\begin{aligned}
[\text{Expressions}] \quad & e ::= o \mid b \mid (\lambda\ (\vec{x})\ e) \mid x \\
& \mid (e\ \vec{e}) \mid (\text{if0}\ e\ e\ e) \\
[\text{Value Literals}] \quad & b ::= 0 \mid -1 \mid 1 \mid \ldots \\
[\text{Primitives}] \quad & o ::= + \mid \text{cons} \mid \text{car} \mid \text{cdr} \mid \ldots \\
[\text{Values}] \quad & v ::= o \mid b \mid (v,v) \mid (\vec{x}, e, \rho) \\
[\text{Standard Answers}] \quad & a ::= v \mid \text{error}^{\text{RT}} \\
[\text{Answers}] \quad & \alpha ::= a \mid \text{error}^{\text{SC}} \\
[\text{Environments}] \quad & \rho = x \to v \\
[\text{Size-change Table}] \quad & m \in v \rightharpoonup v \times \vec{g} \\
[\text{Size-change Graph}] \quad & g \in \mathcal{P}(\mathbb{N} \times r \times \mathbb{N}) \\
[\text{Change}] \quad & r ::= \downarrow \mid \overline{\downarrow}
\end{aligned}
$$

$$
\frac{}{\rho, \bot \vdash e \Downarrow \text{error}^{\text{SC}}} \text{ SC-Err}
\qquad
\frac{}{\rho, m \vdash o \Downarrow o} \text{ SC-Prim}
$$

$$
\frac{}{\rho, m \vdash b \Downarrow b} \text{ SC-Base}
\qquad
\frac{}{\rho, m \vdash (\lambda\ (\vec{x})\ e) \Downarrow (\vec{x}, e, \rho)} \text{ SC-Lam}
$$

$$
\frac{}{\rho, m \vdash x \Downarrow \rho(x)} \text{ SC-Var}
\qquad
\frac{\rho, m \vdash e \Downarrow 0 \qquad \rho, m \vdash e_1 \Downarrow \alpha}{\rho, m \vdash (\text{if0}\ e\ e_1\ e_2) \Downarrow \alpha} \text{ SC-If-T}
$$

$$
\frac{\rho, m \vdash e \Downarrow v \text{ where } v \neq 0 \qquad \rho, m \vdash e_2 \Downarrow \alpha}{\rho, m \vdash (\text{if0}\ e\ e_1\ e_2) \Downarrow \alpha} \text{ SC-If-F}
$$

$$
\frac{\rho, m \vdash e \Downarrow (\vec{x}, e', \rho') \qquad \rho, m \vdash \overrightarrow{e_x} \Downarrow \overrightarrow{v_x} \qquad \rho'[\overrightarrow{x \mapsto v_x}], \boldsymbol{upd}(m, (\vec{x}, e', \rho'), \overrightarrow{v_x}) \vdash e' \Downarrow \alpha}{\rho, m \vdash (e\ \overrightarrow{e_x}) \Downarrow \alpha} \text{ SC-App-Clo}
$$

**Figure 3.** Syntax and semantics of $\lambda_{\text{SCT}}$.

$$
\boldsymbol{upd}\ :\ m \times v \times \vec{v} \to m\ \cup\ \{\bot\}
$$
$$
\boldsymbol{upd}(m, v, \vec{v}_n) = m[v \mapsto (\vec{v}_n, [])], \text{ if } v \notin m
$$
$$
\boldsymbol{upd}(m, v, \vec{v}_n) = \begin{cases} m[v \mapsto (\vec{v}_n, g_n :: \vec{g}_{n-1})] \\ \quad \text{if } \boldsymbol{prog?}(g_n :: \vec{g}_{n-1}) \\ \bot \quad \text{otherwise} \end{cases}
$$
$$
\text{where } (\vec{v}_{n-1}, \vec{g}_{n-1}) \equiv m(v)
$$
$$
\text{and } g_n = \boldsymbol{graph}(\vec{v}_{n-1}, \vec{v}_n)
$$

$$
\boldsymbol{graph}\ :\ \vec{v} \times \vec{v} \to g
$$
$$
\begin{aligned}
\boldsymbol{graph}(\vec{v}, \vec{v}') = & \{(i \downarrow j) \mid v_i \in \vec{v}, v_j \in \vec{v}', v_j \prec v_i\} \\
& \cup\ \{(i \overline{\downarrow} j) \mid v_i \in \vec{v}, v_j \in \vec{v}', v_j = v_i\}
\end{aligned}
$$

$$
(;)\ :\ g \times g \to g
$$
$$
\begin{aligned}
g_0\ ;\ g_1 = & \{(i \downarrow k) \mid (i \downarrow j) \in g_0, (j\ r\ k) \in g_1\} \\
& \cup\ \{(i \downarrow k) \mid (i\ r\ j) \in g_0, (j \downarrow k) \in g_1)\} \\
& \cup\ \{(i \overline{\downarrow} k) \mid (i \overline{\downarrow} j) \in g_0, (j \overline{\downarrow} k) \in g_1, \\
& \qquad \nexists j.(i \downarrow j) \in g_0 \wedge (j\ r\ k) \in g_1, \\
& \qquad \nexists j.(i\ r\ j) \in g_0 \wedge (j \downarrow k) \in g_1\}
\end{aligned}
$$

$$
\boldsymbol{prog?}\ :\ \vec{g} \to \mathbb{B}
$$
$$
\boldsymbol{prog?}(g_n \ldots g_1) = \bigwedge\nolimits_{1 \le i \le j \le n} \boldsymbol{desc?}(g_i; \ldots; g_j)
$$

$$
\boldsymbol{desc?}\ :\ g \to \mathbb{B}
$$
$$
\boldsymbol{desc?}(g) = (g = g; g) \implies \exists i.(i \downarrow i) \in g
$$

**Figure 4.** Updating and monitoring size-change

with the current argument list as well as the empty graph sequence is stored.

Function $\boldsymbol{graph}$ computes a size-change graph from two value lists. For each value $v_j$ at index $j$ in the latter list that is observed to be strictly smaller than some value $v_i$ at index $i$ in the former list, an arc $(i \downarrow j)$ is included in the graph. When the values are equal, we include $(i \overline{\downarrow} j)$ instead.

The composition (;) of two size-change graphs ($g_0$ and $g_1$) includes an arc $(i \downarrow k)$ if there is an arc $(i\ r\ j)$ in $g_0$ and $(j\ r\ k)$ in $g_1$, with at least one arc being a strict descent. If $i$ propagates to $k$ only through non-ascendancy, the weaker arc $(i \overline{\downarrow} k)$ is included.

Finally, predicate $\boldsymbol{prog?}$ checks for the lack of violation to the size-change termination principle: a graph sequence $g_n \ldots g_1$ violates the size-change principle if there exists a sub-sequence $g_i; \ldots; g_j$ (where $1 \le i \le j \le n$) that is both idempotent and lacking of an strict descending arc of a parameter to itself.

### 3.3 Well-founded partial order

Figure 5 shows an example of a well-founded partial order ($\preceq$) on values in $\lambda_{\text{SCT}}$. It is defined on integers by comparing absolute values, and a field of a data-structure is considered smaller than any data-structures that contain it (i.e., latest arguments and size-change graph, potentially returning $\bot$ if there is a size-change violation. If $\boldsymbol{upd}$ does not return a table, the evaluation aborts with an error as in rule *[SC-Err]*.

### 3.2 Updating and monitoring size-change graphs

Figure 4 lists helper functions that update and monitor SCT.

Function $\boldsymbol{upd}$ takes the size-change table ($m$), function ($v$), and its latest arguments ($\vec{v}_n$). It computes a new size-change graph ($g_n$) for the transitions from the previous arguments ($\vec{v}_{n-1}$) to these new arguments, ensures that the new graph sequence ($g_n :: \vec{g}_{n-1}$) does not violate the size-change property, and then updates the graph in $m$. If function $v$ has not been applied before and there is no entry in $m$, a trivial entry

$$\prec, \preceq \subseteq v \times v$$
$$n_1 \prec n_2 \quad \text{if } |n_1| < |n_2|$$
$$v \prec (v', \_) \quad \text{if } v \preceq v'$$
$$v \prec (\_, v') \quad \text{if } v \preceq v'$$
$$v \preceq v' \quad \text{if } v \prec v' \text{ or } v = v'$$

**Figure 5.** Example well-founded partial order $\preceq$

the tail of any list is considered *less than* than the original list). Although simple, this relation is sufficient to check for termination in most programs that descend on integers and data-structures. If a program descends following a different order, the user of $\lambda_{\mathrm{SCT}}$ can replace the default order with an appropriate one.

### 3.4 Totality of evaluation

We may note that all programs in $\lambda_{\mathrm{SCT}}$ terminate, either by adhering to the size-change principle, or by violating it and aborting with an error.

**Theorem 3.1** (Termination of $\lambda_{\mathrm{SCT}}$). *For all $e$, $\rho$, $m$, where $fv(e) \subseteq dom(\rho)$, $\rho, m \vdash e \downarrow \alpha$ for some $\alpha$.*

### 3.5 Soundness and completeness

The size-change property is a safe over-approximation to ensure termination. The correctness of monitoring this property can therefore be understood as any strategy that satisfies the following properties:

**soundness:** if a program evaluates to a value under the modified semantics, running the program without termination checking gives the same result.

**SCT-completeness:** if a program terminates and maintains the under the standard semantics, running that program under the modified size-change property under the standard semantics, running that program under the modified semantics with termination checking gives the same result. program under the modified semantics with termination checking gives the same result.

In addition, because all programs terminate under the modified semantics when termination checking is enabled, all diverging programs are caught as error-raising programs.

We now formally establish the correctness of $\lambda_{\mathrm{SCT}}$'s size-change monitoring semantics with respect to its standard semantics.[1]

**Theorem 3.2** (Soundness of size-change monitoring in $\lambda_{\mathrm{SCT}}$). *If $\rho, m \vdash e \Downarrow a$, then $\rho \vdash e \Downarrow a$.*

*Proof.* By induction on the derivation of $\rho, m \vdash e \Downarrow a$. □

---

CC-BASE
$$\overline{\rho, m \vdash b \Downarrow b, \{m\}}$$

CC-APP
$$\rho, m \vdash e \Downarrow (\vec{x}, e', \rho'), \{m'\ldots\}$$
$$\rho, m \vdash \overline{e_x} \Downarrow \overline{v_x}, \{m_x\ldots\}$$
$$\rho'[\overline{x \mapsto v_x}], ext(m, (\vec{x}, e', \rho'), \overline{v_x}) \vdash e' \Downarrow a, \{m''\ldots\}$$
$$\overline{\rho, m \vdash (e\ \overline{e_x}) \Downarrow a, \{m'\ldots\} \cup \overline{\{m_x\ldots\}} \cup \{m''\ldots\}}$$

$$ext \quad : \quad m \times v \times \vec{v} \to m$$
$$ext(m, v, \vec{v}_n) \quad = \quad m[v \mapsto (\vec{v}_n, [])], \text{if } v \notin m$$
$$ext(m, v, \vec{v}_n) \quad = \quad m[v \mapsto (\vec{v}_n, g_n :: \vec{g}_{n-1})]$$
$$\text{where} \quad (\vec{v}_{n-1}, \vec{g}_{n-1}) \equiv m(v)$$
$$\text{and} \quad g_n = graph(\vec{v}_{n-1}, \vec{v}_n)$$

**Figure 6.** Call-sequence Semantics of $\lambda_{\mathrm{SCT}}$.

**Corollary 3.3** (Size-change monitoring catches divergence). *If program $e$ diverges under the standard semantics, then $\{\}, \{\} \vdash e \Downarrow \mathrm{error}^{\mathrm{SC}}$.*

*Proof.* From Theorem 3.1, $e$ either evaluates to a standard answer or $\mathrm{error}^{\mathrm{SC}}$ under size-change monitoring. By contrapositive of Theorem 3.2, $e$ evaluates to $\mathrm{error}^{\mathrm{SC}}$ if it diverges. □

***A semantics that produces call sequences*** Before stating and proving completeness of size-change monitoring, we define a mostly-standard semantics that also evaluates to set of size-change tables along with the answer, but performs no guarding against any size-change violation. It is in lock-step with the standard semantics, and resembles the terminating semantics in accumulating the size-change table. Figure 6 shows this semantics.

**Lemma 3.4** (Completeness of call-sequence semantics). *If $\rho \vdash e \Downarrow v$ then $\rho, \{\} \vdash e \Downarrow v, \{m\ldots\}$ for some $\{m\ldots\}$.*

*Proof.* By induction on the derivation of $\rho \vdash e \Downarrow v$. □

**Lemma 3.5** (Completeness of size-change monitoring with respect to call-sequence semantics). *If $\rho, m \vdash e \Downarrow \mathrm{error}^{\mathrm{SC}}$ and $\rho, m \vdash e \Downarrow v, \{m'\ldots\}$ then there exists $m_i$ in $\{m'\ldots\}$ and $v$ such that $\neg prog?(g)$ where $(\overline{v_x}, g) = m_i(v)$.*

*Proof.* By induction on the derivation of $\rho, m \vdash e \Downarrow \mathrm{error}^{\mathrm{SC}}$. □

**Theorem 3.6** (Completeness of size-change monitoring in $\lambda_{\mathrm{SCT}}$). *If $\rho, \{\} \vdash e \Downarrow \mathrm{error}^{\mathrm{SC}}$ and $\rho \vdash e \Downarrow v$ then $\rho, \{\} \vdash e \Downarrow v, \{m\ldots\}$ such that there exists $m_i$ in $\{m\ldots\}$ and $v$ such that $\neg prog?(g)$ where $(\overline{v_x}, g) = m_i(v)$.*

*Proof.* Follows from Lemma 3.4 and Lemma 3.5. □

$$[\text{Expressions}] \quad e ::= \dots \mid (\text{term/c } e)$$

$$[\text{Values}] \quad v ::= \dots \mid \text{term/c}(\vec{x}, e, \rho)$$

**Wrap-Lam**
$$\frac{\rho \vdash e \Downarrow (\vec{x}, e, \rho)}{\rho \vdash (\text{term/c } e) \Downarrow \text{term/c}(\vec{x}, e, \rho)}$$

**App-Term**
$$\frac{\rho \vdash e \Downarrow \text{term/c}(\vec{x}, e', \rho') \qquad \rho \vdash \overrightarrow{e_x} \Downarrow \overrightarrow{v_x}}{\rho'[\overrightarrow{x \mapsto v_x}], \; \boldsymbol{upd}(\{\}, (\vec{x}, e', \rho'), \overrightarrow{v_x}) \vdash e' \; \Downarrow \; \alpha}{\rho \vdash (e \; \overrightarrow{e_x}) \Downarrow \alpha}$$

**Figure 7.** Syntax and semantics of $\lambda_{\text{CSCT}}$.

### 3.6 Termination checking as a contract

It can be useful to enforce termination checking selectively on parts of the code rather than on the entire program. We present a simple extension to $\lambda_{\text{SCT}}$ called $\lambda_{\text{CSCT}}$, which adds a construct $(\text{term/c } e)$ that guards $(e)$ with a contract ensuring it behaves as a size-change-terminating function. Other than executing the bodies of contract-guarded functions, the $\lambda_{\text{CSCT}}$ semantics is similar to the standard semantics. Figure 7 shows the key extension to $\lambda_{\text{CSCT}}$'s syntax and semantics. The *[Wrap-Lam]* rule shows the introduction of a termination-checked function. Only closures are capable of violating SCT in $\lambda_{\text{SCT}}$, so we only wrap closures and return other values as-is.

## 4 Static SCT verification

Given termination formulated as a dynamically checkable property, we can systematically turn these dynamic checks into static verification by building on prior work in higher-order symbolic execution [34, 46, 50].

Symbolic execution extends the standard semantics with symbolic values that can stand for any values (including higher-order values), and maintains a path-condition, which is a formula about facts that must hold for symbolic values on each path. Because termination checks ultimately decompose into "less-than" checks, which check for a definite descent of values along a well-founded partial order, there is no special challenge in using symbolic execution for size-change termination checking. Symbolic execution can readily leverage SMT solvers for precise reasoning about path-conditions, proving termination that depends on sophisticated path-sensitivity.

Although symbolic execution has traditionally been used to find bugs [7, 26, 30, 38, 39] as opposed to verifying programs as correct, we can apply a well studied technique for abstracting the operational semantics through finitizing the program's dynamic components [15, 49] and obtain a verification that particular errors cannot occur at run-time.

$$[\text{Values}] \quad v ::= \dots \mid s$$

$$[\text{Symbolic Values}] \quad s ::= x \mid b \mid (o \; \vec{s})$$

$$[\text{Path Conditions}] \quad \phi = \vec{s}$$

**Sym-If-T**
$$\frac{\rho, \phi \vdash e \Downarrow_s s, \phi' \qquad \rho, (= s \; 0) :: \phi' \vdash e_1 \Downarrow_s \alpha, \phi''}{\rho, \phi \vdash (\text{if0 } e \; e_1 \; e_2) \Downarrow_s \alpha, \phi''}$$

**Sym-If-F**
$$\frac{\rho, \phi \vdash e \Downarrow_s s, \phi' \qquad \rho, (\neq s \; 0) :: \phi' \vdash e_2 \Downarrow_s \alpha, \phi''}{\rho, \phi \vdash (\text{if0 } e \; e_1 \; e_2) \Downarrow_s \alpha, \phi''}$$

**Figure 8.** Semantics of symbolic $\lambda_{\text{SSCT}}$.

### 4.1 Extended semantics

Figure 8 shows extension to $\lambda_{\text{SCT}}$, called $\lambda_{\text{SSCT}}$ that allows symbolic execution, as well as the key extension to the semantics that enables symbolic execution.

We extend the set of values $(v)$ with *symbolic values* $(s)$, which can stand for any value. The semantics of $\lambda_{\text{SSCT}}$ must then account for symbolic values, which means some expressions can nondeterministically evaluate to multiple answers to soundly over-approximate all the cases resulting from possible instantiations of symbolic values. Symbolic execution maintains a *path-condition* $(\phi)$ that characterizes each path, which is a set of symbolic values assumed to have evaluated to true (interpreted as a conjunction).

With symbolic values, orders between values are necessarily more conservative. The size-change graphs computed between symbolic value lists in Figure 4 have, in general, no more arcs than in the concrete case. Each arc now represents a must-descend or must-non-ascend relationship over all possible concrete paths. A sufficiently precise symbolic execution, coupled with effective SMT solving, can maintain a graph with enough arcs to prove that functions will always maintain their size-change properties.

**Proposition 4.1** (Soundness of static verification). *If* $\{\} \vdash e \Downarrow v$ *and* $\{\} \vdash e_1 \Downarrow v_1$ *and* $(e \; e_1)$ *diverges, then* $\{\}, \{\} \vdash ((\text{term/c } e) \; s) \Downarrow_s \text{error}^{\text{SC}}, \phi'$ *(s is a fresh symbolic value).*

*Proof.* Follows from soundness of dynamic checking of size-change termination (Theorem 3.2) and soundness of higher-order symbolic execution [33]. □

### 4.2 Ackermann revisited

Now consider again the example ack, a termination-checked Ackermann function shown in Section 2.

Suppose ack's precondition is that its arguments are natural numbers. To verify ack, we apply the function on symbolic natural numbers m and n that have passed ack's precondition with the path-condition $\{(\geq \text{ m 0}), (\geq \text{ n 0})\}$.

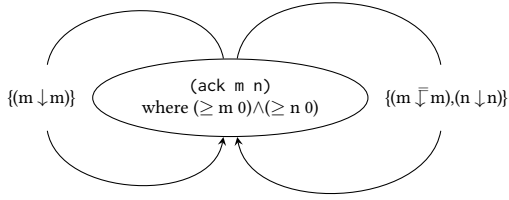

**Figure 9.** Abstract call and size-change graphs for `ack`.

With these symbolic inputs, execution considers all three branches, and accumulates in the path-condition assumptions about the values: in the first branch, `m` is 0; in the second branch, `m` is positive and `n` is 0; in the last branch, both `m` and `n` are positive.

The first branch simply returns and does not trigger any size-change monitoring. The second branch reaches a recursive call with the path-condition {(≥ `m` 0),(≠ `m` 0),(= `n` 0)}. The recursive call proceeds, checking for all relationships that can be established between the old and new arguments as in Figure 4. In this case, with the path-condition that `m` is positive, symbolic execution easily proves that (- `m` 1) is less than `m` according to the partial order defined in Figure 5. No other definite order can be established between the new arguments (- `m` 1), 1 and the old ones `m`, `n`. This gives the new size-change graph of {(m ↓ m)}.[2] We extend `ack`'s set of size-change graphs with this new graph. In addition, symbolic execution can prove the new call to `ack` receives the same path-condition as the previous call: both new arguments (- `m` 1) and 1 are natural numbers.

The third branch reaches the inner recursive call to `ack` before reaching the outer one. The path-condition, again, is sufficient for establish the descent from `n` to (- `n` 1) and maintenance of `m`, yielding the new graph {(m ⊤̲ m),(n ↓ n)}. When execution reaches the outer recursive call to `ack`, the descent from `n` to (- `n` 1) can be straightforwardly established. In each case, symbolic execution can also prove that the new arguments are natural numbers.

Figure 9 summarizes all the ways `ack` can call itself recursively. Because no composition of size-change graphs drawn from this set can yield a graph that violates the size-change principle (i.e. one that is both idempotent and lacking of a self-descent arc), `ack` never violates size-change termination.

## 5 Implementation and evaluation

We implement the semantics presented in Section 3 as a library in the Racket programming language through instrumentation of the application form.

---

[2]We use variable names instead of indices for graph nodes in this section.

An application (`f x ...`) in Racket is syntactic sugar for (`#%app f x ...`), and libraries can modify what an application means by redefining the `#%app` form. For our purposes, we redefine the application form to implement the rules [SC-App-Clo] in Figure 3 [App-Term] in Figure 7. If size-change termination is being enforced, the `#%app` form looks up the size-change table to guard against violations.

We evaluate two techniques to maintain size-change tables. The first technique wraps each application with code that imperatively updates and restores the table. The second uses continuation-marks [10]. The former can be implemented in most languages, and gives relatively good performance, but breaks proper tail calls. The latter is simple to implement in languages with support for continuation marks, and preserves tail calls, but shows high overheads in tight loops.

Our semantics implicitly assumes that closures can be compared structurally for equality, which is not possible in practice. We instead hash the closure and consider all closures with the same hash code to be equivalent. This preserves soundness as the table $m$ cannot grow infinitely, but could produce false positive error reports. Note that this incompleteness does not affect the static analysis, which is derived from the semantics itself. Future work includes runtime support for more precise comparison between closures.

In addition, we expose a parameter specifying the custom partial order for use in termination checks, with a default implementation as described in Figure 5.

Although a naive implementation would be prohibitively expensive, with a few optimizations, the overhead can be brought down to acceptable for the goal of debugging

***Reducing monitoring frequency*** The construction and checking of size-change graphs is expensive, but need not be performed each time a function calls itself recursively. Because strict progress down any well-founded partial order can only be maintained a finite number of times, any non-SCT program will violate the size-change principle regardless of the monitoring frequency. We therefore use exponential backoff to reduce the frequency of extending and monitoring each function's size-change. This significantly reduces the monitoring overhead, although risks keeping data from earlier iterations live for longer necessary.

***Avoiding instrumentation for known functions*** Functions that are known to terminate need no instrumentation. We maintain a white-list of primitives known to terminate.

***Monitoring size-change graphs only for loop entries*** We identify "loop entries" to monitor instead of constructing and monitoring a size-change graph for each function. For example, suppose `even?` and `odd?` are mutual recursive functions, where the top-level context calls `even?`, then only `even?` is a loop-entry and requires size-change monitoring.

**Table 1.** Evaluation on terminating programs

| Program | Dyn. | Static | LH | Isabelle | ACL2 |
|---|---|---|---|---|---|
| sct-1 (rev) | ✓ | ✓ | ✓[R] | ✓ | ✓ |
| sct-2 | ✓ | ✓ | ✗ | ✓[R] | ✓ |
| sct-3 (ack) | ✓ | ✓ | ✓[A] | ✓ | ✓ |
| sct-4 | ✓ | ✓ | ✗ | ✓ | ✓ |
| sct-5 | ✓ | ✓ | ✗ | ✓ | ✓ |
| sct-6 | ✓ | ✓ | ✗ | ✓ | ✓ |
| ho-sc-ack | ✓ | ✗ | _[T] | _[T] | _[H] |
| ho-sct-fg | ✓ | ✓ | ✗ | ✓ | _[H] |
| ho-sct-fold | ✓ | ✓ | ✓[A] | ✓ | _[H] |
| isabelle-perm | ✓ | ✓ | ✗ | ✓ | ✓ |
| isabelle-f | ✓ | ✗ | ✗ | ✓ | ✓ |
| isabelle-foo | ✓ | ✗ | ✗ | ✓ | ✓ |
| isabelle-bar | ✓ | ✗ | ✗ | ✓ | ✓ |
| isabelle-poly | ✓ | ✗ | ✗ | ✗ | ✗ |
| acl2-fig-2 | ✓[O] | ✗ | ✗ | ✗ | ✗ |
| acl2-fig-6 | ✓ | ✗ | ✗ | ✗ | ✗ |
| acl2-fig-7 | ✓ | ✗ | ✗ | ✗ | ✓ |
| lh-gcd | ✓ | ✗ | ✓ | ✓ | ✓ |
| lh-map | ✓ | ✓ | ✓ | ✓ | _[H] |
| lh-merge | ✓ | ✓ | ✓[A] | ✓ | ✓ |
| lh-range | ✓[O] | ✗ | ✓[A] | ✗ | ✓ |
| lh-tfact | ✓ | ✓ | ✓ | ✓ | ✓ |
| dderiv | ✓ | ✓ | | | A: With annotations |
| deriv | ✓ | ✗ | | | O: Custom partial order |
| destruct | ✓ | ✗ | | | H: No H.O. functions |
| div | ✓ | ✓ | | | T: Not typable |
| nfa | ✓ | ✓ | | | R: Rewritten to use |
| scheme | ✓ | ✗ | | | pattern matching |

## 5.1 Evaluation

We evaluate the effectiveness and efficiency of size-change monitoring. Effective monitoring should allow all or most terminating programs to finish execution, and quickly catch diverging programs. Efficient monitoring should introduce little overhead compared to execution without monitoring.

### 5.1.1 Effectiveness and efficiency on terminating programs

Table 1 shows terminating programs we use to evaluate the dynamic checks and static analysis of terminating contracts. The programs were collected from previous work on termination checking: size-change termination for first-order programs (sct) [29]; size-change termination for higher-order programs (ho-sct) [41]; Liquid Haskell (lh) [52]; Isabelle [28]; ACL2 [31]; and a collection of larger Scheme benchmarks that terminate by the size-change principle.

The table shows the precision of dynamic checking and static analysis, as well as comparison with other systems where possible. Most programs are small and under 15 lines.

The largest program is scheme with 1,100 lines, which implements an interpreter for R5RS Scheme that interprets the mergesort algorithm on a list of strings. We did our best efforts to translate programs from one system to another. For example, sct-2 is originally an untyped program composing a heterogeneous list which cannot be typed in Liquid Haskell and Isabelle. We translated sct-2 to work with an equivalent custom tree data-type.

Several cases where the programs need modifications to be successfully verified by the systems are annotated in the table. For example, sct-1 and sct-2 originally use conditionals, and can only be verified when converted to use pattern-matching. Some other programs are only verified successfully with annotational help on termination, such as explicit lexical ordering (e.g. lh-merge), or a custom partial-order (e.g. acl2-fig-2). Some programs are not expressible in all systems. For example, ACL2 cannot check higher-order programs, and the type systems in Liquid Haskell and Isabelle do not support the Y-combinator that has self-applications (e.g. ho-sct-ack). To our surprise, current versions of the tools cannot check some of their own benchmarks despite our best efforts to reproduce (e.g. isabelle-poly for Isabelle; acl2-fig-2 and acl2-fig-6 for ACL2). Overall, our system works well for a wide range of programs and idioms, including higher-order untyped programs with moderate side effects (such as in the Scheme benchmarks).

Figure 10 shows the slowdown of dynamic checks for select programs: factorial, sum, and merge-sort, as well as their interpreted version inside a Scheme interpreter. These programs demonstrate that different patterns of computation incur different amounts of overhead from size-change monitoring. For programs that do significant work between recursive calls, such as factorial or the Scheme interpreter, overhead is negligible. For programs that don't do significant work between recursive calls, such as sum, the overhead is significant. For programs that operate over large data-structures such as merge-sort, overhead is much more significant. That the overhead stays fixed when the input grows (for continuation-mark implementation on tight loops, approximately two orders of magnitude) suggests that further optimization effort to trim down the constant factor can make monitoring suitable for realistic uses.

### 5.1.2 Effectiveness on diverging programs

We also evaluate dynamic monitoring on non-terminating programs to determine how quickly the monitoring system catches divergence. These programs include modified versions of correct programs, as well as one originally incorrect program (nfa) that our static analysis discovered. Because violation of the size-change principle tend to show up in early iterations, our dynamic contracts catch the error very early, resulting in immeasurable delay from the start of the program to the point where divergence is detected.

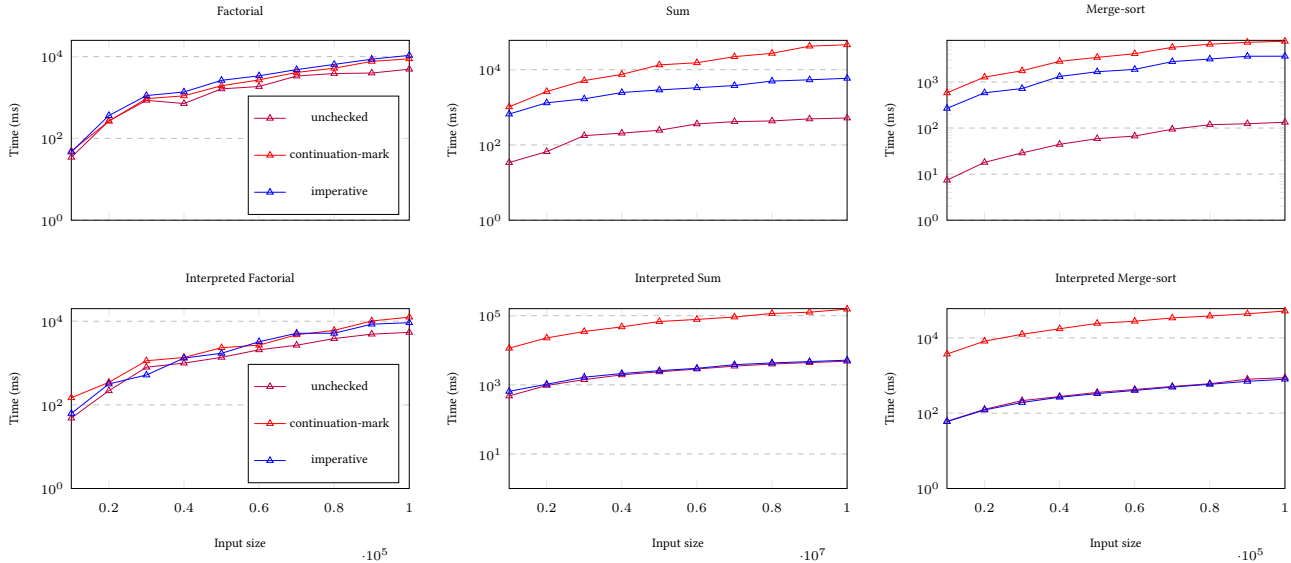

**Figure 10.** Slow-down of monitoring `factorial`, `sum`, and `merge-sort`, and the Scheme interpreter running them

The `nfa` program is particularly interesting, because it is a Scheme benchmark that has been around for decades. It is a program that implements a non-deterministic finite automaton of the regular expression `((a|c)*bcd)|(a*bc)`, then run the automaton on the string $a^{133}bc$. The following function implements one state recognizing the sub-expression `(a|c)*` with the bug underlined:

```
1  (define (state1 input)
2    (and (not (null? input))
3         (or (and (char=? (car input) #\a)
4                  (state1 (cdr input)))
5             (and (char=? (car input) #\c)
6                  (state1 input))
7             (state2 input))))
```

The bug was never discovered, because the particular benchmark input did not trigger the divergence, and most static analysis only check for partial correctness. Our static analysis was the first to discover this error after many years.

## 6 Related work

Our work builds on the size-change termination (SCT) approach [29] and on approaches to static contract verification via symbolic execution [33, 34]. We relate our contributions to dynamic and static termination checking, and then to static contract verification.

### 6.1 Dynamic termination checking

To the best of our knowledge, no existing work enforces termination dynamically using behavioral contracts. Related work has investigated dynamic loop detection, nontermination auditing, and more restricted declarative languages.

The auditing tool Looper [6] dynamically monitors a Java program in order to detect nontermination using concolic (concrete and symbolic) execution. Along the path of a potentially nonterminating loop, it derives a path condition paired with a memory map (an encoding of heap values at the end of a loop iteration as a function of their initial values), and uses an SMT solver to check if the initial path condition (after zero iterations) implies itself under the loop iteration's memory map. If this fails, Looper will observe another iteration and record a new path condition and memory map. When each path condition implies the next (under that iteration's memory map), in a cyclic chain that terminates with the original path condition, the program will not terminate.

Unlike our contracts, Looper does not monitor code for nontermination during normal execution; instead, it is deployed by an auditor to determine whether an apparent loop is an actual one. While Looper can provide an affirmative proof that code will not terminate, our approach will signal that a function does not obey SCT, a more conservative notion of termination. This means our approach is susceptible to false positives and may blame functions which do always terminate, but will never permit nontermination. Looper, on the other hand, is susceptible to false negatives and may fail to prove an execution to be definitively nonterminating. Looper's soundness is also contingent on all changes to memory being visible and accounted for in the memory map, which is not always the case in C due to external state and shared-memory parallelism.

Jolt [8] (and successor Bolt [27]) is an infinite-loop detection and recovery tool for C programs. It instruments C

code to dynamically monitor for loops that are in the exact same state at two consecutive iterations. Compared with LOOPER, this is an especially conservative detection for nontermination, however JOLT also has a facility for skipping the program counter past the end of the loop to recover from nontermination and show that this simple technique is effective in many cases (sometimes depending on inputs).

There are also dynamic termination schemes for more restricted languages. For example, dynamic checking for active database rules [3], or queries in general logic programs [12, 42]. Shen et al. [42] exploits features unique to SLDNF-trees to identify loop goals with a provably finite term-size. Codish and Taboch [12] provides a declarative fixed-point semantics that captures termination properties (for an interpretation of Prolog) with the explicit goal of facilitating the extraction of a static analysis using abstract interpretation.

## 6.2 Static termination checking

A variety of approaches have been used for static verification of termination and nontermination. None of these systems combine dynamic and static verification in a single system, or allow terminating and nonterminating components to be composed. We begin with the systems we compare with in §5.1.

Jones and Bohr [25] extend the SCT approach to higher-order languages—specifically, the untyped $\lambda$-calculus. As all values in this language are functions, they select the "height" of a closure as its size. Sereni and Jones [41] then extended this approach to handle user-defined datatypes and general recursion. This work was not empirically evaluated in the context of a real programming system [40], but establishes techniques we build on. SCT has been extended to monotonicity constraints, which have been shown to be more general than classical SCT [11]; these could be formulated as a dynamic contract in future work.

Manolios and Vroon [31] develops a static analysis for automatic termination proofs in the context of the ACL2 system—a functional language and first-order logic for theorem proving. All programs admitted by ACL2 must be terminating, as nontermination could render it inconsistent, however manual termination proofs are complex and require deep expertise. The paper's approach uses precise calling-context graphs in order to refine static control flow with path feasibility based on accumulating governors (sets of branch points governing control flow for a subexpression). Strongly connected components are then further refined using a calling-context measure in order to discover a well-founded order over which parameters descend. A major innovation on traditional SCT approaches is the refinement of feasible paths using governors. Our approach analogously tracks path conditions for static verification. Their method was effective at proving more than 98% of the roughly 10k functions of the ACL2 regression suite terminating. Krauss

[28] then extends the approach to Isabelle/HOL and certifies the termination proofs with LCF-style theorem proving.

LIQUIDHASKELL uses termination proving to ensure precision and soundness for its refinement type system in the presence of lazy evaluation [51]. Subtle unsoundness can result from using refinement types in conjunction with call-by-name evaluation and the direct approach to fixing this unsoundness, by expressing potential nontermination as a type refinement, leads to substantial imprecision. LIQUID-HASKELL bridges this gap by encoding size-change invariants, over user-specified well-founded metrics, directly into the existing type system (as further type refinements). This permits proofs over programs to circularly depend on termination proofs during SMT solving. Broadly this same approach is taken to directly encode termination proofs, via size-change refinements, with dependent types in DEPENDENTML [54]. LIQUIDHASKELL has a scalable implementation, used to verify correctness and termination properties over a corpus of real-world Haskell libraries ($\geq 10k$ LOC). TEA is also a termination analysis for Haskell, based on techniques of path analysis and abstract reduction [35].

TNT is a concolic executor for statically enumerating nonterminating *lassos* in C programs—paths that fold back on themselves, forming a nonterminating loop [23]. Unlike dynamic approaches such as ours, or that of LOOPER, TNT is not statically precise enough to handle cases that rely on symbolic shape information such as cyclic lists.

Velroyen and Rümmer [53] use a modal logic allowing predicates to be written that are qualified by a program expression they pertain to. Qualified formulae are trivially rendered true by a diverging program, so a manifest contradiction (i.e., false) being interpreted as true constitutes a proof of nontermination for the qualifying expression. The approach then uses a refinement process to identify the specific conditions on data that will lead to proving this contradiction. This system was only evaluated on small expressions ($\leq 25$ lines) in a language of pure built-in expressions, assignments, conditionals, and while loops.

APROVE is a system for automating termination (and nontermination) proofs of term-rewriting systems (TRSs) Giesl et al. [18, 19, 22] built using the dependency pair framework [2, 20]. Unlike previous methods for proving TRSs terminating, which required the right-hand side of each rewrite rule to be simplified compared with its left-hand side, the dependency pair framework only requires corresponding subterms at recursive calls to be simplified. This innovation is analogous to the SCT approach's requirement that arguments be descending over some well-founded order as opposed to static control-flow being strictly stratified. Giesl et al. [21] extends the dependency pair framework to higher-order functions. Thiemann and Giesl [45] contrast and synthesize the dependency pair framework with SCT.

Numerous techniques have been proposed and evaluated for verifying termination in languages such as C and Java,

where higher-order programming is uncommon. TERMINATOR [13, 14, 36] and transition invariants [24, 37, 47] as well as others [1, 4, 5, 44] have seen extensive development, and share some key ideas with our approach, but differ substantially in goals and language from our system, and thus make significantly different choice in approach.

TERMINATOR is a program analysis and verification tool for proving termination of C programs statically, which has been used to prove the termination of low-level programs such as Windows device drivers. Like our system, it relies on the "indirect approach" described in the introduction—find a safety property which implies termination, add a check for that property to the program, and verify using an existing tool that the check cannot fail. The key difference with our approach is in the choice of property. TERMINATOR aims to prove termination of tricky first-order loops, and thus must find potentially-complex custom ranking functions (found via Podelski and Rybalchenko [36]) for each program to be verified. To find these properties, it relies on a counter-example guided abstraction refinement (CEGAR) [9] loop which attempts to verify termination using an off-the-shelf verifier and refines the property upon failure. Terminator starts with a very simple property and repeatedly improves it, generating complex predicates with non-trivial relationships between multiple variables. In contrast, our approach (similar to other approaches for higher-order languages) picks a *single* general safety property and uses it for all programs. This limits the ability of our tool to verify the termination of loops such as those TERMINATOR aims at, but allows our tool to run as a contract without first requiring several static runs of a static verifier. Additionally, constructing static verification tools for heap-manipulating imperative programs is much trickier in the higher-order setting we consider.

### 6.3 Soft contract verification

Our static termination checking relies on the ability to go from an operational semantics with dynamic enforcement to a sound static analyzer—a capability we take from a series of results on static contract checking by Nguyễn et al. [33, 34]. This work showed that sound higher-order symbolic execution could be used to provide contract-based soft verification and counter-example generation for rich languages including user-defined data structures and contracts as well as higher-order functions and state. We re-use this work by retargeting it to contracts that enforce size-change termination, but otherwise retain the central ideas; it is a goal of our work that it composes with existing contract systems.

## 7 Conclusion

Termination is a fundamental program correctness property, but uncheckable even at runtime. To avoid this limitation, we adapt the size-change principle from static termination analysis to perform dynamic checking of termination, exploiting the insight that every infinite execution must have a call that fails to follow the size change principle. This leads to the first run-time mechanism for enforcing termination in a general-purpose programming system. As it is formulated as a behavioral contract, this also makes it the first contract for total correctness. By checking termination as a contract, we can enforce termination in settings where static checking is fundamentally impossible, as in an interpreter.

Further, we compose our dynamic checking strategy with prior work showing how to statically verify compliance with contracts in higher-order languages to produce a novel static checker for program termination—without any termination-specific work. We compare our static checker against three state-of-the-art custom tools on their own benchmarks, and find that ours is able to statically verify programs that exceed the capacities of each of the existing tools.

Sound dynamic enforcement of liveness properties opens up new possibilities for program correctness, analysis, and specification—in this paper we have taken only the first step.

## Acknowledgments

We are grateful to Amir Ben-Amram, Michael W. Hicks, and Éric Tanter for comments on early drafts of this work. This work is supported by part by the National Science Foundation awards #1846350 and #1763922.

$$\frac{}{\rho \vdash o \Downarrow o} \text{ Prim} \qquad \frac{}{\rho \vdash b \Downarrow b} \text{ Base}$$

$$\frac{}{\rho \vdash (\lambda\ (\vec{x})\ e) \Downarrow (\vec{x}, e, \rho)} \text{ Lam} \qquad \frac{}{\rho \vdash x \Downarrow \rho(x)} \text{ Var}$$

$$\text{App-Clo} \quad \frac{\rho \vdash e \Downarrow (\vec{x}, e', \rho') \qquad \rho \vdash \overrightarrow{e_x} \Downarrow \overrightarrow{v_x} \qquad \rho'[\overrightarrow{x \mapsto v_x}] \vdash e' \Downarrow \alpha}{\rho \vdash (e\ \overrightarrow{e_x}) \Downarrow \alpha}$$

$$\frac{}{\rho, \bot \vdash e \Downarrow \mathsf{error}^{SC}} \text{ SC-Err} \qquad \frac{}{\rho, m \vdash o \Downarrow o} \text{ SC-Prim}$$

$$\frac{}{\rho, m \vdash b \Downarrow b} \text{ SC-Base} \qquad \frac{}{\rho, m \vdash (\lambda\ (\vec{x})\ e) \Downarrow (\vec{x}, e, \rho)} \text{ SC-Lam}$$

$$\frac{}{\rho, m \vdash x \Downarrow \rho(x)} \text{ SC-Var} \qquad \text{SC-If-T} \quad \frac{\rho, m \vdash e \Downarrow 0 \qquad \rho, m \vdash e_1 \Downarrow \alpha}{\rho, m \vdash (\mathsf{if0}\ e\ e_1\ e_2) \Downarrow \alpha}$$

$$\text{SC-If-F} \quad \frac{\rho, m \vdash e \Downarrow v \text{ where } v \neq 0 \qquad \rho, m \vdash e_2 \Downarrow \alpha}{\rho, m \vdash (\mathsf{if0}\ e\ e_1\ e_2) \Downarrow \alpha}$$

$$\text{SC-App-Clo} \quad \frac{\rho, m \vdash e \Downarrow (\vec{x}, e', \rho') \qquad \rho, m \vdash \overrightarrow{e_x} \Downarrow \overrightarrow{v_x} \qquad \rho'[\overrightarrow{x \mapsto v_x}], \mathit{update}(m, (\vec{x}, e', \rho'), \overrightarrow{v_x}) \vdash e' \Downarrow a}{\rho, m \vdash (e\ \overrightarrow{e_x}) \Downarrow a}$$

**Figure 11.** Standard and Terminating semantics of $\lambda_{\mathrm{SCT}}$.

| | | |
|---|---|---|
| [Expressions] | $e ::= o \mid b \mid (\lambda\ (\vec{x})\ e) \mid x$ | |
| | $\quad \mid (e\ \vec{e}) \mid (\mathsf{term}/\mathsf{c}\ e)$ | |
| [Value Literals] | $b ::= 0 \mid -1 \mid 1 \mid \ldots$ | |
| [Primitives] | $o ::= + \mid \mathsf{cons} \mid \mathsf{car} \mid \mathsf{cdr} \mid \ldots$ | |
| [Values] | $v ::= o \mid b \mid (v, v) \mid (\vec{x}, e, \rho)$ | |
| | $\quad \mid \mathsf{term}/\mathsf{c}(\vec{x}, e, \rho)$ | |
| [Size-change Table] | $m \in v \rightharpoonup \vec{v} \times g$ | |
| [Size-change Graph] | $g \in \mathcal{P}(\mathbb{N} \times r \times \mathbb{N})$ | |
| [Change] | $r ::= \Downarrow \mid \overline{\Downarrow}$ | |

**Figure 12.** Syntax of $\lambda_{\mathrm{CSCT}}$.

## A  Supplemental material

### Proof of Theorem 3.1

*Proof.* Consider an infinite sequence of function calls. By Lemma A.1 below, there's a closure that keeps being called. The sequence of arguments to this closure cannot satisfy the size-change property an infinite number of times. The diverging program that results in this call sequence will be killed. □

**Lemma A.1** (Recurring closure). *Along any infinite sequence of function calls, there is at least one closure that is called infinitely often.*

*Proof.* Consider the sequence of closures

$$(x_1, e_1, \rho_1), \ldots, (x_i, e_i, \rho_i), \ldots$$

along the infinite call sequence.

- Case 1: The closures come from a finite set. At least one must repeat infinitely often.
- Case 2: There are fresh closures that keep being generated dynamically. Because new infinite closures must be generated through finite $\lambda$ forms, there must be some infinite subset of closures $(x_i, e_i, \_)$ generated by one same form $(\lambda\ (x_i)\ e_i)$. Let $(x_m, e_m, \_)$ be

$$\frac{}{\rho \vdash o \Downarrow o} \textsc{Prim} \qquad \frac{}{\rho \vdash b \Downarrow b} \textsc{Base} \qquad \frac{}{\rho \vdash (\lambda\ (\vec{x})\ e) \Downarrow (\vec{x}, e, \rho)} \textsc{Lam} \qquad \frac{}{\rho \vdash x \Downarrow \rho(x)} \textsc{Var} \qquad \frac{\rho \vdash e \Downarrow (\vec{x}, e, \rho)}{\rho \vdash (\text{term}/\text{c}\ e) \Downarrow \text{term}/\text{c}(\vec{x}, e, \rho)} \textsc{Wrap-Lam}$$

$$\frac{\rho \vdash e \Downarrow o}{\rho \vdash (\text{term}/\text{c}\ e) \Downarrow o} \textsc{Wrap-Prim} \qquad \frac{\rho \vdash e \Downarrow (\vec{x}, e', \rho') \qquad \rho \vdash \vec{e_x} \Downarrow \vec{v_x}}{\rho'[\overrightarrow{x \mapsto v_x}] \vdash e' \Downarrow \alpha}{\rho \vdash (e\ \vec{e_x}) \Downarrow \alpha} \textsc{App-Clo} \qquad \frac{\rho \vdash e \Downarrow \text{term}/\text{c}(\vec{x}, e', \rho') \qquad \rho \vdash \overrightarrow{\rho_x} \Downarrow \vec{v_x}}{\rho'[\overrightarrow{x \mapsto v_x}], \boldsymbol{update}(\{\}, (\vec{x}, e', \rho'), \vec{v_x}) \vdash e' \boldsymbol{\Downarrow} \alpha}{\rho \vdash (e\ \vec{e_x}) \Downarrow \alpha} \textsc{App-Term}$$

$$\frac{}{\rho, \bot \vdash e \boldsymbol{\Downarrow} \text{error}^{\textsc{SC}}} \textsc{SC-Err} \qquad \frac{}{\rho, m \vdash o \boldsymbol{\Downarrow} o} \textsc{SC-Prim} \qquad \frac{}{\rho, m \vdash b \boldsymbol{\Downarrow} b} \textsc{SC-Base} \qquad \frac{}{\rho, m \vdash (\lambda\ (\vec{x})\ e) \boldsymbol{\Downarrow} (\vec{x}, e, \rho)} \textsc{SC-Lam} \qquad \frac{}{\rho, m \vdash x \boldsymbol{\Downarrow} \rho(x)} \textsc{SC-Var}$$

$$\frac{\rho, m \vdash e \boldsymbol{\Downarrow} (\vec{x}, e, \rho)}{\rho, m \vdash (\text{term}/\text{c}\ e) \boldsymbol{\Downarrow} \text{term}/\text{c}(\vec{x}, e, \rho)} \textsc{SC-Wrap-Lam} \qquad \frac{\rho, m \vdash e \boldsymbol{\Downarrow} o}{\rho, m \vdash (\text{term}/\text{c}\ e) \boldsymbol{\Downarrow} o} \textsc{SC-Wrap-Prim} \qquad \frac{\rho, m \vdash e \boldsymbol{\Downarrow} (\vec{x}, e', \rho') \qquad \rho, m \vdash \vec{e_x} \boldsymbol{\Downarrow} \vec{v_x}}{\rho'[\overrightarrow{x \mapsto v_x}], \boldsymbol{update}(m, (\vec{x}, e', \rho'), \vec{v_x}) \vdash e' \boldsymbol{\Downarrow} \alpha}{\rho, m \vdash (e\ \vec{e_x}) \boldsymbol{\Downarrow} \alpha} \textsc{SC-App-Clo}$$

$$\frac{\rho, m \vdash e \boldsymbol{\Downarrow} \text{term}/\text{c}(\vec{x}, e', \rho') \qquad \rho, m \vdash \vec{e_x} \boldsymbol{\Downarrow} \vec{v_x}}{\rho'[\overrightarrow{x \mapsto v_x}], \boldsymbol{update}(m, (\vec{x}, e', \rho'), \vec{v_x}) \vdash e' \boldsymbol{\Downarrow} \alpha}{\rho, m \vdash (e\ \vec{e_x}) \boldsymbol{\Downarrow} \alpha} \textsc{SC-App-Term}$$

**Figure 13.** Semantics of $\lambda_{\textsc{CSCT}}$.

$$\frac{}{\rho, m \vdash o \Downarrow o, \{m\}} \textsc{CC-Prim} \qquad \frac{}{\rho, m \vdash b \Downarrow b, \{m\}} \textsc{CC-Base}$$

$$\frac{}{\rho, m \vdash (\lambda\ (\vec{x})\ e) \Downarrow (\vec{x}, e, \rho), \{m\}} \textsc{CC-Lam}$$

$$\frac{\rho, m \vdash e \Downarrow (\vec{x}, e', \rho'), \{m'\ ...\}}{\rho, m \vdash \vec{e_x} \Downarrow \vec{v_x}, \{m_x\ ...\}}{\rho'[\overrightarrow{x \mapsto v_x}], \boldsymbol{ext}(m, (\vec{x}, e', \rho'), \vec{v_x}) \vdash e' \Downarrow a, \{m''\ ...\}}{\rho, m \vdash (e\ \vec{e_x}) \Downarrow a, \{m'\ ...\} \cup \{m_x\ ...\} \cup \{m''\ ...\}} \textsc{CC-App}$$

**Figure 14.** Call-sequence Semantics of $\lambda_{\textsc{CSCT}}$.

the set of closures whose body $e_m$ contains this form $(\lambda\ (x_i)\ e_i)$.

– Claim: There must be some closure $(x_j, e_j, \rho_j)$ that keeps being called infinitely often.
– Proof: By induction on the lexical depth of the term $(\lambda\ (x_m)\ e_m)$.
  * Subcase 1: $(\lambda\ (x_m)\ e_m)$ has lexical depth 0 (i.e. it is a top-level $\lambda$). Because it is not enclosed by any $\lambda$, the closure $(x_m, e_m, \{\})$ is created only once. By assumption, $(x_m, e_m, \{\})$ is called infinitely often to dynamically create the infinite closure set $(x_i, e_i, \_)$.
  * Subcase 2: $(\lambda\ (x_m)\ e_m)$ is directly enclosed by $(\lambda\ (x_n)\ e_n)$.
    · Subsubcase 2a: The set $(x_m, e_m, \_)$ is finite: at least one of them is called infinitely often to generate the infinite closure set $(x_i, e_i, \_)$.
    · Subsubcase 2b: The set $(x_m, e_m, \_)$ is infinite: apply the induction hypothesis on $(\lambda\ (x_n)\ e_n)$ (where new $i$ is $m$ and new $m$ is $n$).                    □